\documentclass[12pt]{iopart}
\usepackage{epsfig}
\usepackage{citesort}

\newcommand{\gtrsim}{\,\rlap{\lower3.7pt\hbox{$\mathchar\sim$}}
\raise1pt\hbox{$>$}\,}
\newcommand{\lesssim}{\,\rlap{\lower3.7pt\hbox{$\mathchar\sim$}}
\raise1pt\hbox{$<$}\,}

\long\def\dump#1{}

\def\a{\alpha}

\def\k{\kappa}

\def\m{\mu}

\def\t{\tau}

\def\D{\Delta}

\def\G{\Gamma}

\def\O{\Omega}

\def\ve{\varepsilon}

\def\mb{\overline{m}}
\begin{document}

\title{Quantum Zeno effect and the impact of flavor in leptogenesis}

\author{S.~Blanchet, P.~Di Bari and G.~G.~Raffelt}
\address{Max-Planck-Institut f\"ur Physik
(Werner-Heisenberg-Institut)\\
F\"ohringer Ring 6, 80805 M\"unchen, Germany\\
~\\
MPP-2006-155 (17 November 2006)}

\begin{abstract}
In thermal leptogenesis, the cosmic matter-antimatter asymmetry is
produced by {\it CP\/} violation in the decays $N\to\ell+\Phi$ of
heavy right-handed Majorana neutrinos $N$ into ordinary leptons
$\ell$ and Higgs particles $\Phi$. If some charged-lepton Yukawa
couplings are in equilibrium during the leptogenesis epoch, the
$\ell$ interactions with the background medium are flavor sensitive
and the coherence of their flavor content defined by $N\to\ell+\Phi$
is destroyed, modifying the efficiency of the inverse decays. We
point out, however, that it is not enough that the flavor-sensitive
processes are fast on the cosmic expansion time scale, they must be
fast relative to the $N\leftrightarrow\ell+\Phi$ reactions lest the
flavor amplitudes of $\ell$ remain frozen by the repeated
$N\leftrightarrow\ell+\Phi$ ``measurements''. Our more restrictive
requirement is significant in the most interesting ``strong wash-out
case'' where $N\leftrightarrow\ell+\Phi$ is fast relative to the
cosmic expansion rate. We derive conditions for the unflavored
treatment to be adequate and for flavor effects to be maximal. In
this ``fully flavored regime'' a neutrino mass bound
survives. To decide if this bound can be circumvented in the
intermediate case, a full quantum kinetic treatment is required.
\end{abstract}

%\maketitle

%%%%%%%%%%%%%%%%%%%%%%%%%%%%%%%%%%%%%%%%%%%%%%%%%%%%%%%%%%%%%%%%%%%%%%
\section{Introduction}                        \label{sec:introduction}
%%%%%%%%%%%%%%%%%%%%%%%%%%%%%%%%%%%%%%%%%%%%%%%%%%%%%%%%%%%%%%%%%%%%%%

The see-saw mechanism~\cite{Minkowski:1977sc,Yanagida:1979as,%
Gell-Mann:1980vs,Glashow:1979nm,Barbieri:1979ag,Mohapatra:1979ia} is
an elegant way to understand neutrino masses and mixing and at the
same time, with leptogenesis~\cite{Fukugita:1986hr}, provides a
natural mechanism for generating the matter-antimatter asymmetry of
the Universe by virtue of {\it CP\/}-violating decays
$N\to\ell+\Phi$ of heavy right-handed Majorana neutrinos $N$ into
ordinary leptons $\ell$ and Higgs particles $\Phi$. In comparison
with other baryogenesis scenarios, leptogenesis has the unique
advantage of relying on an ingredient of Physics Beyond the Standard
Model, neutrino masses, that is experimentally established.
Moreover, the neutrino mass values suggested by flavor oscillation
experiments are optimal in the sense that leptogenesis would operate
in a mildly ``strong wash-out regime''~\cite{Buchmuller:2003gz}.
This means that the inverse processes $\ell+\Phi\to N$ are efficient
enough to wash out all contributions to the final asymmetry that
depend on initial conditions, but not too efficient to prevent
successful leptogenesis. This is true for hierarchical light
neutrino schemes, while in the case of quasi-degenerate neutrinos,
leptogenesis provides a stringent upper bound on neutrino masses,
$m_1\leq 0.1\,{\rm eV}$, holding when a hierarchical right-handed
neutrino spectrum is assumed~\cite{Buchmuller:2003gz}.

Traditionally, flavor effects were neglected in leptogenesis, i.e.,
the lepton asymmetry produced by $N\to\ell+\Phi$ and the wash-out
effects caused by the inverse processes were treated as if $\ell$
had no flavor properties. Recently it was recognized, however, that
important modifications arise if some of the charged-lepton Yukawa
interactions are in
equilibrium~\cite{Barbieri,Endoh:2003mz,Nardi:2006fx,Abada:2006fw}.
In particular, there is an additional source of {\it CP\/} violation
that can make the single-flavored {\it CP\/} asymmetries larger than
the total. This additional source includes a dependence on
low-energy phases, in contrast to unflavored leptogenesis
\cite{Endoh:2003mz,Nardi:2006fx}. In fact, successful leptogenesis
stemming only from Majorana and Dirac phases is possible
\cite{Abada:2006ea,Blanchet:2006be,Pascoli:2006ie,Branco:2006ce}. A
second consequence is a reduction of the wash-out efficiency,
generally by different amounts in each flavor. These modifications
together contribute to enhance the final asymmetry, an effect that
increases for a larger neutrino mass scale. Therefore, flavor
effects can relax the neutrino mass bound of the unflavored
scenario~\cite{Abada:2006fw}.

The purpose of our present study is to explore the conditions for
flavor effects to be relevant in leptogenesis and if the neutrino
mass bound can indeed be circumvented. The past literature is
unclear on this point in the following sense. It was stressed that
the $\ell$ interactions with the background medium are flavor
sensitive and the coherence of the $\ell$ flavor content defined by
$N\to\ell+\Phi$ can be destroyed if some of the charged-lepton
Yukawa couplings are in thermal equilibrium during the leptogenesis
epoch. On the other hand, the reactions $N\leftrightarrow\ell+\Phi$
constantly re-generate the $\ell$ flavor composition and if these
reactions are fast, $\ell$ is frozen in its coherent flavor state by
the quantum Zeno effect~\cite{Stodolsky:1986dx,Raffelt:1992uj}. In
other words, the charged-lepton Yukawa couplings and the
$N\leftrightarrow\ell+\Phi$ reactions generally single out different
directions in flavor space. The $\ell$ density matrix in flavor
space will dominantly stay in the direction favored by the fastest
term of the kinetic equation. Therefore, we expect flavor effects to
be important if the ``flavor measurements'' by the background medium
are fast relative to the $N\leftrightarrow\ell+\Phi$ reactions, a
condition that was also stated in Ref.~\cite{Nardi:2006fx}. On the
other hand, the actually used requirement in the previous literature
was the weaker condition that the flavor-sensitive reactions are
fast on the cosmic expansion scale. Of course, at the end of the
leptogenesis epoch when the $N\leftrightarrow\ell+\Phi$ reactions no
longer track equilibrium, this weaker condition is correct. Our main
concern is that previous treatments may not properly cover the
important strong wash-out case, where $N\leftrightarrow\ell+\Phi$ is
fast relative to the expansion rate for most of the leptogenesis
epoch.

In order to identify the conditions for flavor effects to be
relevant for the final baryon asymmetry we split the problem into
two parts. First, in Section~\ref{sec:unflavored}, we derive a
condition for the unflavored treatment to be adequate. Next, in
Section~\ref{sec:maxflavored}, we obtain a condition for flavor
effects to be maximally effective (``fully flavored regime''). In
these cases two different sets of classical Boltzmann equations
apply to calculate the final asymmetry. In
Section~\ref{sec:transition} we consider the neutrino mass bound to
hold in the unflavored case and show that it cannot be circumvented
in the fully flavored case. However, there is an intermediate regime
between the fully flavored and unflavored cases where a full quantum
kinetic equation is required to decide on the neutrino mass bound.
In Section~\ref{sec:beyond} we briefly discuss the limitations of
our simple rate comparison. We summarize our findings in
Section~\ref{sec:conclusions}.

%%%%%%%%%%%%%%%%%%%%%%%%%%%%%%%%%%%%%%%%%%%%%%%%%%%%%%%%%%%%%%%%%%%%%%
\section{Unflavored case}                       \label{sec:unflavored}
%%%%%%%%%%%%%%%%%%%%%%%%%%%%%%%%%%%%%%%%%%%%%%%%%%%%%%%%%%%%%%%%%%%%%%

\subsection{Unflavored leptogenesis}

In order to derive a condition for the unflavored leptogenesis
treatment to be adequate, we first retrace the steps leading to the
final asymmetry prediction. Adding to the Standard Model three
right-handed (RH) neutrinos with a Majorana mass term $M$ and Yukawa
couplings $h$, after spontaneous symmetry breaking a Dirac mass term,
$m_{\rm D}=v\,h$, is generated by the vev $v$ of the Higgs boson. In the
see-saw limit, $M\gg m_{\rm D}$, the spectrum of neutrino masses
splits into two sets, a very heavy one, $M_3\geq M_2 \geq M_1$,
almost coinciding with the eigenvalues of $M$, and a light one,
$m_3\geq m_2\geq m_1$, corresponding to the eigenvalues of the light
neutrino mass matrix. It is given by the see-saw
\hbox{formula~\cite{Minkowski:1977sc,Yanagida:1979as,%
Gell-Mann:1980vs,Glashow:1979nm,Barbieri:1979ag,Mohapatra:1979ia}},
\begin{equation}
m_{\nu}= - m_{\rm D} {1\over M}\, m_{\rm D}^T \,.
\end{equation}
Neutrino oscillation experiments measure two light neutrino mass
squared differences. In a normal scheme one has
$m^{\,2}_3-m_2^{\,2}=\Delta m^2_{\rm atm}$ and
$m^{\,2}_3-m_1^{\,2}=\Delta m^2_{\rm sol}$, whereas in an inverted
scheme one has $m^{\,2}_3-m_2^{\,2}=\Delta m^2_{\rm sol}$ and
$m^{\,2}_2-m_1^{\,2}=\Delta m^2_{\rm atm}$. For $m_1\gg m_{\rm atm}
\equiv \sqrt{\Delta m^2_{\rm atm}+\Delta m^2_{\rm sol}}$ the
spectrum is quasi-degenerate, while for $m_1\ll m_{\rm sol}\equiv
\sqrt{\Delta m^2_{\rm sol}}$ it is fully hierarchical.

In the early Universe, the decays of the RH neutrinos into leptons
and Higgs bosons generally produce a net lepton number that is
partly converted into a net baryon number by sphaleron processes if
the temperature exceeds about $100\,{\rm GeV}$. We will follow
common practice and assume that the final asymmetry is determined
only by the decays of the lightest RH neutrino, although a scenario
where the asymmetry is produced by the next-to-lightest states is
also possible~\cite{DiBari:2005st,vives}.

For now we also assume that the
flavor composition of the leptons produced in the decays does not
affect the final asymmetry (unflavored limit). Neglecting thermal
effects in $\D L=1$ scatterings~\cite{Giudice:2003jh} and spectator
processes~\cite{Buchmuller:2001sr,Nardi:2005hs}, the final asymmetry
is determined by a simple set of two kinetic equations, one for the
RH neutrino abundance and one for the $B-L$
asymmetry~\cite{Buchmuller:2004nz},
\begin{eqnarray}\label{unflke1}
 {dN_{N_1}\over dz} & = & -D\,\left(N_{N_1}-N_{N_1}^{\rm eq}\right)
 \\\label{unflke2}
 {dN_{B-L}\over dz} & = &
 \ve_{1}\,D\,\left(N_{N_1}-N_{N_1}^{\rm eq}\right)-
     \left(W_1^{\rm ID}+\D W\right)\,N_{B-L} \, ,
\end{eqnarray}
where $z\equiv M_1/T$. With $N_{B-L}$ and $N_{N_1}$ we denote the
abundances per RH neutrino $N_1$, taken to be in ultra-relativistic
thermal equilibrium. The actual $N_1$ equilibrium abundance is
$N_{N_1}^{\rm eq}=z^2\,{\cal K}_2(z)/2$, where ${\cal K}_i(z)$ are
the modified Bessel functions.

Introducing the decay parameter $K_1\equiv
\widetilde{\G}_{1}/H_{T=M_1}$, defined as the ratio of the total
decay width to the expansion rate at $T=M_1$, the decay term can be
written as
\begin{equation}
 D \equiv {\G_{\rm D}\over H\,z}=K_1\,z\,
 \left\langle\gamma_1^{-1}\right\rangle   \, ,
\end{equation}
where $\langle\gamma_1^{-1}\rangle={\cal K}_1(z)/{\cal K}_2(z)$ is
the thermally averaged Lorentz dilation factor. The expansion rate
is
\begin{equation}
 H\simeq \sqrt{8\,\pi^3\,g_{\star}\over 90}\,
 {M_1^2\over M_{\rm Pl}}\,{1\over z^{2}}
 \simeq 1.66\,\sqrt{g_{\star}}\,
 {M_1^2\over M_{\rm Pl}}\,{1\over z^{2}}\,
\end{equation}
and $g_{\star}=g_{\rm SM}=106.75$ is the total number of thermally
excited degrees of freedom.

The wash-out term caused by inverse decays is, after subtracting the
resonant $\D L=2$ processes,
\begin{equation}\label{WID}
 W_1^{\rm ID}(z) \equiv {1\over 2}\,{\Gamma_1^{\rm ID}(z)\over H(z)\,z}=
 {1\over 4}\,K_1\,{\cal K}_1(z)\,z^3 \, ,
\end{equation}
where $\Gamma_1^{\rm ID}=\G_{\rm D}\,N_{N_1}^{\rm eq}/N_{\ell}^{\rm
eq}$ is the inverse decay rate. The non-resonant  $\Delta L=2$
contribution dominates in the non-relativistic regime and is
approximately
\begin{equation}\label{DWmax}
 \Delta W(z) \simeq  {w\over z^2}
 \left({M_1\over 10^{10}\,{\rm GeV}}\right)
 \left({\mb\over {\rm eV}}\right)^2 \;,
\end{equation}
where
\begin{equation}
w={9\sqrt{5}\,M_{\rm Pl}\,10^{-8}\,{\rm GeV}^3 \over
4\pi^{9/2}\,g_\ell\,\sqrt{g_{\star}}\,v^4}\simeq 0.186
\end{equation}
and $\mb\equiv \sqrt{\sum_i\,m_i^2}$.

The evolution of $N_{B-L}$ is explicitly, denoting initial
quantities with the subscript ``in,''
\begin{equation}\label{NBmL}
 N_{B-L}(z)=N_{B-L}^{\rm in}\,
 \exp\left\lbrace -\int_{z_{\rm in}}^z dz'\,
 \left[W_1^{\rm ID}(z')+\D W(z')\right]\right\rbrace
 +\ve_{1}\,\k_{1}(z)
\end{equation}
with the efficiency factor
\begin{equation}\label{ef}
 \k_{1}(z)\equiv -\int_{z_{\rm in}}^z dz'\,{dN_{N_1}\over dz'}\,
 \exp\left[-\int_{z'}^z dz''\,[W_1^{\rm ID}(z'')+\D W(z'')]\right]\,.
\end{equation}
An approximate analytic expression for the final efficiency factor
at the end of the leptogenesis epoch and valid in the strong wash-out
regime ($K_1\gg 1$) is
\begin{equation}
 \k_1^{\rm f}(m_1,M_1,K_1)\simeq \k(K_1)\,
 \exp\left[-{w\over z_B(K_1)}\,
 \left(M_1\over 10^{10}\,{\rm GeV}\right)\,
 \left(\bar{m}\over {\rm eV}\right)^2\right]
\end{equation}
where,
\begin{equation}\label{k1a}
 \k(K_1) \simeq
 {2\over K_{1}\,z_{B}(K_1)}\,
 \left(1-e^{-K_{1}\,z_{B}(K_1)/2}\right) \, .
\end{equation}
Further,
\begin{equation}
 z_B(K_1)\simeq 2+4\,K_1^{0.13}\,e^{-2.5/K_1}
\end{equation}
is an approximate expression for that value of $z$ where $W_1^{\rm
ID}(z_B)\simeq 1$, i.e., where the wash-out term from inverse decays
becomes ineffective~\cite{Blanchet:2006dq}. In the weak wash-out regime
the expression for the final efficiency factor depends on the value
of $N_{N_1}^{\rm in}$. For an initial thermal equilibrium abundance
the given expression still holds. For an initial vanishing abundance the
final efficiency factor is the sum of two contributions
with opposite sign and approximate analytic expressions can be found
in~\cite{Buchmuller:2004nz}. The weak wash out, acting less efficiently
on the positive contribution, prevents a full cancellation.

Assuming either that the initial asymmetry is negligible or that it
is efficiently washed out, the final $B-L$ asymmetry is
$N_{B-L}^{\rm f}\simeq \ve_1\,\k_1^{\rm f}$. Assuming further a
standard thermal history of the universe and accounting for the
sphaleron conversion coefficient $a_{\rm sph}\sim 1/3$, the final
baryon-to-photon ratio at recombination (rec) is
\begin{equation}\label{etaB}
 \eta_B=a_{\rm sph}\,{N_{B-L}^{\rm f}\over N_{\gamma}^{\rm rec}}
 \simeq 0.96\times 10^{-2}\,\ve_1\,\k_1^{\rm f} \,.
\end{equation}
This is the number to be compared with the measured
value~\cite{Spergel:2006hy}
\begin{equation}\label{etaBobs}
 \eta_B^{\rm CMB} = (6.1 \pm 0.2)\times 10^{-10} \, .
\end{equation}

The wash-out rate by inverse decays reaches a maximum $W_1^{\rm
ID}(z_{\rm max})\simeq 0.3\,K_1$ at $z_{\rm max}\simeq 2.4$. In
the weak wash-out regime, when $K_1\lesssim 3.3$, one has
$W_1^{\rm ID}(z)< 1$ for any value of $z$. In this case the wash
out is negligible and the final asymmetry depends on the initial
conditions. In the strong wash-out regime, when $K_1\gtrsim 3.3$,
there is an interval $[z_{\rm on},z_{\rm off}]$ where $W_{1}^{\rm
ID}\geq 1$. The asymmetry produced at $z\lesssim z_{\rm off}$ is
very efficiently washed out and thus the final asymmetry is
essentially what is produced around $z_B\simeq z_{\rm off}$ by
out-of-equilibrium decays of the residual RH neutrinos whose
number corresponds approximately to the final value of the
efficiency factor.

\begin{figure}[ht]
\centerline{\psfig{file=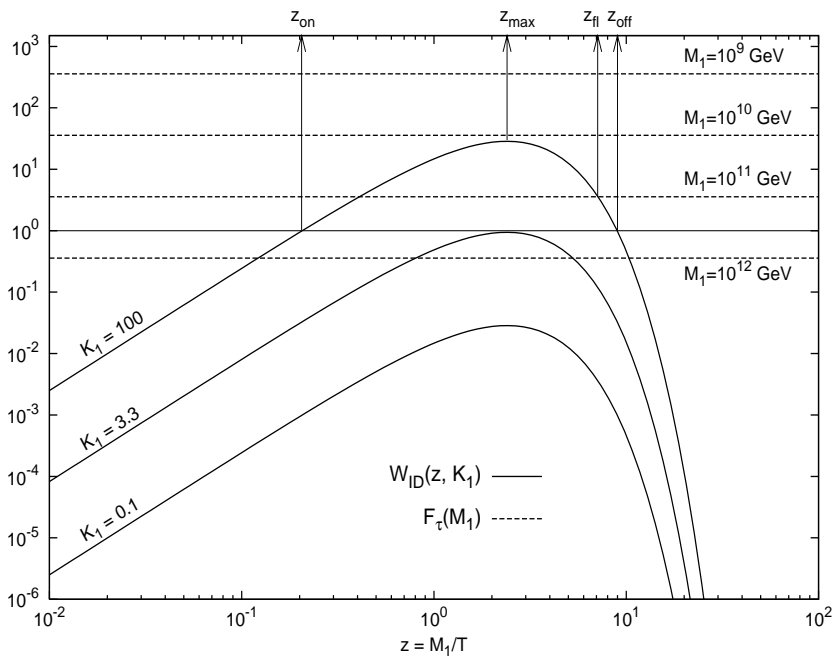,height=12cm,width=9cm,angle=-90}}
\caption{Comparison between the wash-out term $W_1^{\rm ID}(z)$
(thick solid lines), defined in Eq.~(\ref{WID}) and plotted for the
three indicated values of $K_1$, and the charged lepton Yukawa
interaction term $F_{\tau}$, defined in Eq.~(\ref{Ftau}) and plotted
for the indicated values of $M_1$.}\label{fig:washoutterm}
\end{figure}

In Fig.~\ref{fig:washoutterm} we show the wash-out term from inverse
decays for three different values of $K_1$. For $K_1=100$ we show
the interval $[z_{\rm on},z_{\rm off}]$. The  maximum value
$W_1^{\rm ID}(z_{\rm max})\simeq 33$ is reached at $z_{\rm
max}\simeq 2.4$. For $K_1\simeq 3.3$ one has $z_{\rm on}\simeq
z_{\rm max}\simeq z_{\rm off}$. This can be taken as the threshold
value distinguishing between the strong and the weak wash-out
regimes. For $K_1=10^{-1}$ one has $W_1^{\rm ID}\ll 1$ for any value
of $z$. Notice however that even in this case the weak wash out can
be important for successful leptogenesis if the initial abundance
vanishes, since it prevents a full cancellation between two
different sign contributions to the final
asymmetry~\cite{Buchmuller:2004nz}.

\subsection{When are flavor effects important?}

We now turn to the question when flavor effects will modify these
results. The crucial effect is caused by charged-lepton Yukawa
interactions~\cite{Barbieri:1999ma} that occur with a
rate~\cite{Campbell:1992jd} $\Gamma_{\alpha}\simeq 5\times
10^{-3}\,T\,h^2_{\alpha}$. The largest one is for $\a=\t$ where
\begin{equation}
 {\G_{\t}\over H}\simeq \left(10^{12} {\rm GeV}\over T\right)\,.
\end{equation}
Therefore, if $T\gtrsim 10^{12}\,{\rm GeV}$, charged-lepton Yukawa
interactions  are not effective and all processes in the early
Universe are flavor blind, justifying the unflavored treatment. For
$T \lesssim 10^{12}\,{\rm GeV}$, the $\tau$ Yukawa couplings are
strong enough that the scatterings $\tau_L\,\bar{\tau}_R\rightarrow
\Phi^{\dagger}$ are in equilibrium. However, as stressed in the
introduction, this condition is not necessarily sufficient for
important flavor effects to occur because we need to compare the
speed of the Yukawa interactions with that of the RH neutrino decays
and inverse decays. To this end we study the weak and strong
wash-out regimes separately and consider only a two-flavor case
because the $\tau$ lepton Yukawa coupling causes the main
modification.

In the weak wash-out regime, assuming a vanishing initial abundance,
the production of RH neutrinos through inverse decays occurs around
$T\sim M_1$. At this epoch, inverse decays are by definition slower
than the expansion rate. Therefore, the condition $T\lesssim
10^{12}\,{\rm GeV}$ is sufficient to conclude that the
charged-lepton Yukawa interactions are faster than the inverse decay
rate. This translates into the condition $M_1\lesssim 10^{12}\,{\rm
GeV}$ because the RH neutrino production occurs at $T\sim M_1$, in
agreement with the previous
literature~\cite{Nardi:2006fx,Abada:2006fw}.

However, this condition does not guarantee that flavor effects
indeed have an impact on the final asymmetry, because this impact
depends on wash out playing some role. For a vanishing initial
abundance this is the case in that wash-out effects prevent a full
sign cancelation between the asymmetry produced when
$N_{N_1}<N_{N_1}^{\rm eq}$ and the asymmetry produced later on. On
the other hand, for a thermal initial abundance, no such effect
arises from the weak wash out and flavor effects do not modify the
final asymmetry. In any case, one can say that in the limit
$K_1\rightarrow 0$ flavor cannot have effects on the final asymmetry
for any initial abundance. We will come back to this point.

In the strong wash-out regime the situation is very different. The
rate of RH neutrino inverse decays at $T\sim  M_1$ is larger than
the expansion rate. Therefore, we need to compare the charged-lepton
Yukawa rate $\Gamma_{\t}$ with the RH neutrino inverse decay rate
$\Gamma^{\rm ID}_1$. For the unflavored treatment to be valid for
$z\lesssim z_{\rm fl}\leq z_B$ then requires
\begin{equation}\label{M1}
 M_1\gtrsim {10^{12}\,{\rm GeV}\over 2\,W_1^{\rm ID}(z_{\rm fl})}\,,
\end{equation}
where $z_{\rm fl}$ is that value of $z$ where the two rates are
equal. This condition guarantees that at temperatures $T>T_{\rm
fl}=M_1/z_{\rm fl}$ flavor effects will not be able to break the
coherent propagation of lepton states. The final asymmetry is
dominantly produced around $z\sim z_B$. Therefore, the condition for
flavor effects to be negligible is
\begin{equation}\label{unflavored}
M_1\gtrsim 5\times 10^{11}\,{\rm GeV} \, ,
\end{equation}
similar to the weak wash-out regime. However, the
corresponding condition on the temperature
\begin{equation}
 T \gtrsim {10^{12}\,{\rm GeV}\over 2\,z_B(K_1)}
\end{equation}
is now less restrictive.

If one starts with a non-vanishing initial abundance, then the final
asymmetry is also determined by how efficiently the initial value is
washed out; this is described by the integral in Eq.~(\ref{NBmL}).
In this case even a value of $z_{\rm fl}< z_{B}$ would be important
to determine the final asymmetry since wash out in the unflavored
regime would be effective for $z\lesssim z_{\rm fl}< z_{B}$, while a
reduced wash out would apply in the flavored regime at lower
temperatures for $z_{\rm fl}\lesssim z\lesssim z_{B}$.

We conclude that the condition Eq.~(\ref{M1}) obeys the intuitive
expectation that there is always a threshold value for $K_1$ above
which the unflavored case is recovered. In this case the temperature
below which flavor effects play a role indeed becomes smaller and
smaller. The situation is illustrated in Fig.~1 where we compare
$W_{\rm ID}$ with
\begin{equation}\label{Ftau}
F_{\t}\equiv {1\over 2}\,{\G_{\t}\over H\,z}
\simeq {5\times 10^{11}\,{\rm GeV}\over M_1} \, ,
\end{equation}
the analogous quantity for the charged lepton Yukawa interactions.
For any value of $M_1$ and $K_1$, there is a value $z_{\rm fl}$ such
that $F_{\t}\gtrsim W_{\rm ID}$ for $z>z_{\rm fl}$. If $M_1\lesssim
2\times 10^{12}\,{\rm GeV}/K_1$ and $K_1\gtrsim 3.3$, corresponding
to $F_{\tau}\gtrsim W_{\rm ID}(z_{\rm max})$ in the strong wash-out
regime, then $z_{\rm fl}=0$, meaning that flavor effects are
important during the entire thermal history. On the other hand, for
a fixed value of $M_1$, one has $z_{\rm fl}\rightarrow \infty$ for
$K_1\rightarrow \infty$, implying that flavor effects tend to
disappear for sufficiently large values of $K_1$. Notice however
that if $M_1 \lesssim 5\times 10^{11}\,{\rm GeV}$, then $z_{\rm
fl}\gtrsim z_{\rm off}\simeq z_B$ for any value of $K_1$. This
confirms that only for $M_1\gtrsim 5\times 10^{11}\,{\rm GeV} $
flavor effects can be neglected and the unflavored regime is
recovered.

%%%%%%%%%%%%%%%%%%%%%%%%%%%%%%%%%%%%%%%%%%%%%%%%%%%%%%%%%%%%%%%%%%%%%%
\section{Maximum flavor effects}               \label{sec:maxflavored}
%%%%%%%%%%%%%%%%%%%%%%%%%%%%%%%%%%%%%%%%%%%%%%%%%%%%%%%%%%%%%%%%%%%%%%

We now turn to the opposite extreme case when flavor effects are
maximal, the ``fully flavored regime.'' In other words, the
charged-lepton Yukawa interactions are now taken to be so fast that
the lepton flavor content produced in $N\to\ell+\Phi$ on average
fully collapses before the inverse reaction can take place, i.e.,
the $\ell$ density matrix in flavor space is to be taken diagonal in
the charged-lepton Yukawa basis. In this case each single flavor
asymmetry has to be calculated separately because generally the wash
out by inverse decays is different for each flavor. Moreover, the
single-flavored {\it CP\/} asymmetries now have an additional
contribution compared to the total~\cite{Nardi:2006fx,Abada:2006fw}.
Finally, the inverse decay involving a lepton in the flavor $\alpha$
does not wash out as much asymmetry as the one produced by one RH
neutrino decay. The reduction is quantified by the probability
$P^0_{i\a}$, averaged over leptons and anti-leptons, that the lepton
$\ell_i$ produced in the decay of $N_i$ collapses into the flavor
eigenstate $\ell_{\alpha}$. The relevant Boltzmann equations become
\begin{eqnarray}\label{flke}
  {dN_{N_1}\over dz} & = &
  -D_1\,\left(N_{N_1}-N_{N_1}^{\rm eq}\right)
  \nonumber\\
 {dN_{\D_{\a}}\over dz} & = &
 \ve_{1\a}\,D_1\,\left(N_{N_1}-N_{N_1}^{\rm eq}\right)
 -P_{1\a}^{0}\,W_1^{\rm ID}\,N_{\D_{\a}} \, .
\end{eqnarray}
Since we are dealing with the two-flavor case, here $\a=\t$ or $\m$
where the latter stands for a suitable superposition of the $\m$ and
$e$ flavor. Notice also that we defined $\D_{\a}\equiv B/2-L_{\a}$
and therefore the total asymmetry is given by
$N_{B-L}=N_{\D_{\m}}+N_{\D_\t}$.

As in the unflavored case, we next identify the condition for the
fully flavored approximation to hold. The final asymmetry in the
flavor $\a$ is dominantly produced at $z\simeq z_{B\a}\equiv
z_B(K_{1\a})$, where $K_{1\a}\equiv P^0_{1\a}\,K_1$. Therefore, one
must require that $\G_{\a}\gtrsim \G_1^{\rm ID}$ holds already at
$z\sim z_{B\a}$ lest the wash-out reduction takes place too late. We
stress that flavor effects modify the final asymmetry only if the
flavor projection takes place before the wash out by inverse decays
freezes out. Otherwise the wash-out epoch is over and the unflavored
behavior is recovered. It is easy to verify that if the projectors
are set to unity and the equations are summed over flavors, the
kinetic equation for $N_{B-L}$ holding in the unflavored regime is
recovered [cf.~Eq.~(\ref{unflke2})]. Therefore, we require
\begin{equation}\label{M1fl}
 M_1\lesssim {10^{12}\,{\rm GeV}\over 2\,W_1^{\rm ID}(z_{B\a})}
\end{equation}
as an approximate condition for the fully flavored behavior.

\begin{figure}[ht]
\centerline{\psfig{file=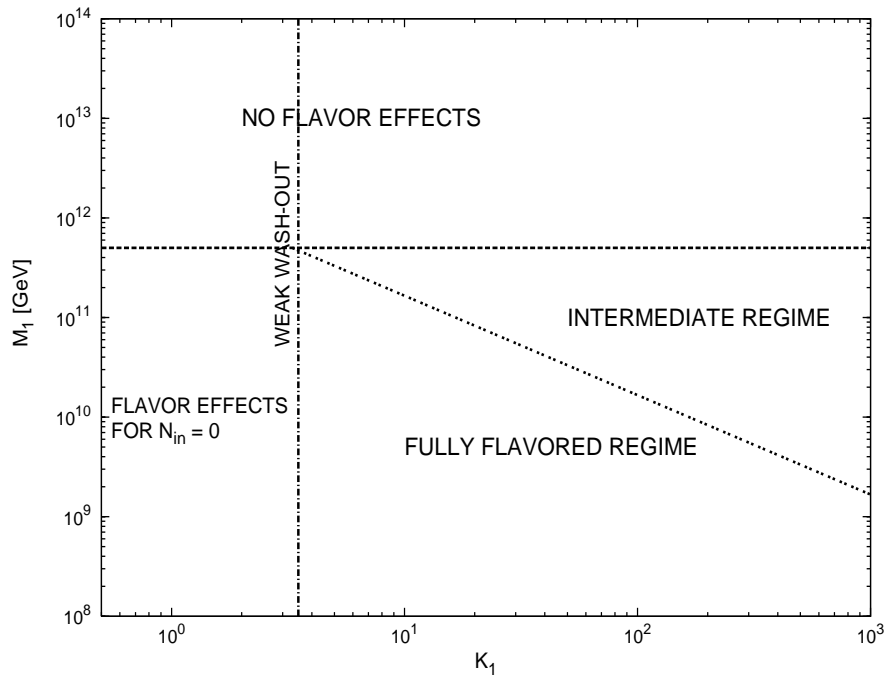,height=12cm,width=9cm,angle=-90}}
\caption{Relevance of flavor effects in schematic regions of
parameters $K_1$ and $M_1$. The region above the
horizontal dashed line corresponds to the condition
(\ref{unflavored}) for $z_{B\a}=z_{\rm max}$.
The vertical dot-dashed line is the border
between the weak and the strong wash-out regime. The region below
the inclined dotted line corresponds to the condition
(\ref{M1fl}) for $z_{B\a}=z_{\rm max}$.} \label{fig:flavorornot}
\end{figure}

In Fig.~\ref{fig:flavorornot} we summarize the different possible
cases in the plane of parameters $K_1$ and $M_1$. For $M_1\gtrsim
5\times 10^{11}\,{\rm GeV}$, above the dashed line, flavor
effects are not important independently of~$K_1$. The condition
Eq.~(\ref{M1fl}), in the most restrictive case when $z_{B\a}=z_{\rm
max}$ and $W_1^{\rm ID}\simeq 0.3\,K_1$, is satisfied below the
inclined dotted line. This case typically occurs in a one flavor
dominated scenario, as we explain below. The vertical dot-dashed line is the
border that separates the weak from the strong wash-out regime in the
unflavored case. In the flavored case the condition $K_1\lesssim 3.3$
still implies the weak wash-out regime because flavor
effects can only reduce the wash out. However, the
condition for the strong wash-out regime can be more
restrictive than $K_1\gtrsim 3.3$, as discussed in
\cite{Blanchet:2006be}.
For $K_1\lesssim 3.3$, flavor effects
modify the final asymmetry only marginally and more
specifically only if the initial abundance vanishes, as indicated in
Fig.~\ref{fig:flavorornot}. On the other hand, for $K_1\gtrsim
3.3$ and below the diagonal line, flavor modifications of the final
asymmetry can be large, especially in the one-flavor dominated
scenario.

There is a region in parameter space where neither condition
Eq.~(\ref{M1}) nor~(\ref{M1fl}) holds. This intermediate regime can
become very large in the case of a one-flavor dominated scenario,
where an order of magnitude enhancement of the final asymmetry is
possible. In this case one of the two projectors is very small
compared to the other and so the wash out is very asymmetric in the
two flavors. On the other hand, if the two-flavored {\it CP\/}
asymmetries are comparable, then the final asymmetry is dominantly
produced into one flavor and deviations from the unflavored regime
can become very large. This scenario is realized, in particular,
when the absolute neutrino mass scale increases, relaxing the
traditional neutrino mass bound.

However, the condition Eq.~(\ref{M1fl}) strongly restricts the
applicability of the one-flavor dominated scenario. Even though the
reduction of the wash out is driven by $K_{1\a}\ll K_1$, implying
$z_{B\a}\ll z_B$, the possibility for flavor effects to be relevant
relies on the dominance of the charged-lepton Yukawa interaction
rate compared to the RH neutrino inverse decay rate that however is
still driven by $K_1$. Therefore, increasing $K_1$, one can enhance
the asymmetry in the one-flavor dominated scenario compared to the
unflavored case, if smaller and smaller values of the projector
$P^0_{1\a}$ are possible. On the other hand, the inverse decay rate
increases so that the fully flavored behavior may no longer apply.
In particular notice that the maximum enhancement of the asymmetry is
obtained when $K_1\gg K_{1\a}\simeq 1$, when
$z_{B\a}\simeq z_{\rm max}$ and the condition (\ref{M1fl}) is
maximally restrictive.

%%%%%%%%%%%%%%%%%%%%%%%%%%%%%%%%%%%%%%%%%%%%%%%%%%%%%%%%%%%%%%%%%%%%%%
\section{Neutrino mass bound}
\label{sec:transition}
%%%%%%%%%%%%%%%%%%%%%%%%%%%%%%%%%%%%%%%%%%%%%%%%%%%%%%%%%%%%%%%%%%%%%%

\begin{figure}[b]
\centerline{\psfig{file=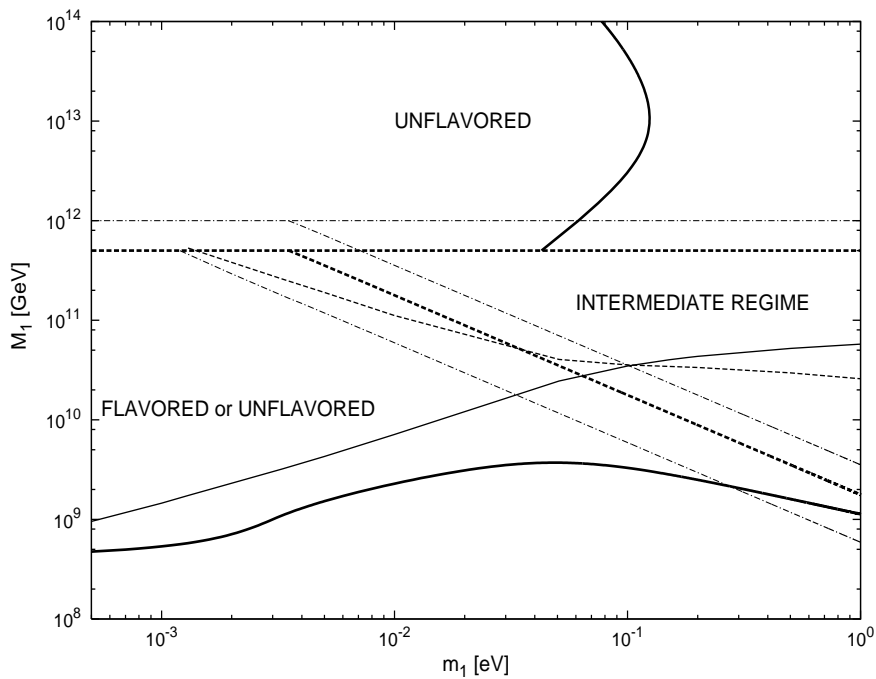,height=12cm,width=9cm,angle=-90}}
\caption{Relevance of flavor effects similar to
Fig.~\ref{fig:flavorornot}, now mapped to schematic regions of
parameters $m_1$ and $M_1$. The region above the horizontal dashed
line corresponds to the condition Eq.~(\ref{unflavored}) for the
applicability of the unflavored regime. The region below the
inclined thick dashed line corresponds to the condition Eq.~(\ref{M1fl})
calculated for that value of $z_{B\a}$ that maximizes the final
asymmetry in the one-flavor dominated scenario and for
$K_1=m_1/m_{\star}$. In this same case, the lower
inclined dot-dashed lines includes also the effect of scatterings
in the condition Eq.~(\ref{M1fl}) while the upper inclined and the horizontal
dot-dashed lines include the effect of oscillations.
The area between the two inclined dot-dashed lines gives an estimation
of the uncertainty on the condition for the
fully flavored regime to hold. The area between the horizontal
thin dot-dashed line and the horizontal thick dashed line
gives an estimation of the uncertainty on the condition
for the unflavored regime to hold. The two
thick solid lines borders the region where successful
leptogenesis is possible: on the left in the unflavored regime
and above in the fully flavored regime. The thin solid line
is a more restrictive border obtained for a specific choice
of the see-saw orthogonal matrix and the thin dashed line is
the corresponding condition Eq.~(\ref{M1fl}). In this case
one has $K_1> m_1/m_{\star}$.
}\label{fig:mMplot}
\end{figure}

One possible consequence of flavor effects is to relax the
traditional upper bound on the neutrino mass that is implied by
successful leptogenesis. In order to explore the impact of our
modified criteria we first recall the origin of this bound in the
unflavored case. Maximizing the final value of the asymmetry over
all see-saw parameters except $M_1$ and $m_1$
yields~\cite{Buchmuller:2004nz}
\begin{eqnarray}\hspace{-18mm}
 {\eta_B^{\rm max}(M_1,m_1)\over \eta_B^{\rm CMB}}&\simeq&
 3.8\,\left({m_{\star}\over m_1}\right)^{1.2}\,
 \left({M_1\over 10^{10}\,{\rm GeV}}\right)\,
 {m_{\rm atm}\over m_1+m_3}\,
 \exp\left[-{w\over z_B}\,{M_1\over 10^{10}\,{\rm GeV}}\,
 \left(\bar{m}\over{\rm eV}\right)^2\right]
 \nonumber\\
 &\geq&1\,,
\end{eqnarray}
where we have approximated $\k(K_1)\simeq 0.5\,K_1^{-1.2}$ \cite{proc}
and we have neglected the dependence of $z_B$ on $K_1$ in the derivative.
This constraint translates into $m_1<m_1^{\rm max}(M_1)$ shown by
the curved solid line in the upper part of Fig.~\ref{fig:mMplot}
where the unflavored behavior obtains. This curve sports an absolute
maximum, $m_1\lesssim 0.12\,{\rm eV}$, for
$M_1\simeq 10^{13}\,{\rm GeV}$.

The possibility that flavor effects could relax this bound is based
on the observation that the flavored {\it CP\/} asymmetries, for
$m_1\gtrsim m_{\rm atm}$, are proportional to $m_1$. On the other
hand, the total asymmetry is suppressed like $m_1^{-1}$,
contributing to the upper bound in the unflavored regime. However,
if $m_1$ increases, then $K_1$ has to increase also so that it is
not guaranteed that the fully flavored treatment remains justified.

Quantitatively, the value of $K_1$ is bounded
from below by \cite{Fujii:2002jw}
\begin{equation}
 K_1\geq {m_1\over m_{\star}} \, ,
\end{equation}
where
\begin{equation}
 m_{\star}={16\, \pi^{5/2}\,\sqrt{g_*} \over 3\,\sqrt{5}}\,
 {v^2 \over M_{\rm Pl}} \simeq 1.08\times 10^{-3}\,{\rm eV}\;.
\end{equation}
For $K_1\geq m_1/m_{\star}\gg 1$, the final asymmetry is maximized
when a one-flavor dominated scenario is realized. In this case the
final asymmetry is approximately $N_{B-L}^{\rm f} \simeq
\ve_{1\a}\,\k_{1\a}^{\rm f}$. The bound on the single flavored {\it
CP\/} asymmetry is~\cite{Abada:2006ea}
\begin{equation}\label{bound}
 |\ve_{1\a}|< \overline{\ve}(M_1)\,\sqrt{P_{1\a}^0}\,
 {m_3\over m_{\rm atm}}\, ,
\end{equation}
where
\begin{equation}
\overline{\ve}(M_1)\equiv {3\over 16\pi}\,{M_1\,m_{\rm atm}\over v^2}
\simeq 10^{-6}\,\left({M_1\over 10^{10}\,{\rm GeV}}\right)
\,\left({m_{\rm atm}\over 0.05\,{\rm eV}}\right) \, .
\end{equation}
It is then possible to find the value of $P_{1\a}^0$ that maximizes
the asymmetry as a function of $K_1$ and the corresponding value of
$z_{B\a}$. Imposing $\eta_B^{\rm max}\geq \eta_B^{\rm CMB}$ implies
a lower bound on $M_1$ as a function of $K_1$.

This limit can be translated into a lower bound on $M_1$ as a
function of $m_1$ by replacing $K_1$ with its minimum value
$m_1/m_{\star}$. In this way the wash out is always minimized and
the final efficiency factor and the final asymmetry are maximized.
Notice that the single-flavored {\it CP\/} asymmetries, like the
total, vanish for $K_1=m_1/m_{\star}$. Therefore, this lower bound
cannot be saturated. Notice moreover that for $m_1\lesssim
m_{\star}$ the one-flavor dominated scenario does not necessarily
holds because it is possible that $K_1\lesssim 1$. Actually for
$K_1\rightarrow 0$, flavor effects disappear and one has to recover
the usual asymptotic value of the lower bound obtained in the
unflavored case for thermal initial abundance, $M_1\gtrsim 4\times
10^{8}\,{\rm GeV}$. For intermediate values of $K_1$ one can use a
simple interpolation. The final result is shown in the bottom part
of Fig.~\ref{fig:mMplot} as a thick solid line.

In Fig.~\ref{fig:mMplot} we also show the condition
Eq.~(\ref{M1fl}), calculated for the same value of $z_{B\a}$ that
maximizes the final asymmetry , but replacing $K_1$
with its minimum value $m_1/m_{\star}$ (thick dashed line).
Since $W_1^{\rm ID}$
increases with $K_1$, this produces a necessary, but not sufficient,
condition in the $m_1$-$M_1$ plane for the fully flavored behavior.
This condition matches the validity of the unflavored regime at
$m_1\simeq 3\times 10^{-3}\,{\rm eV}$ and the lower bound on $M_1$
at $m_1\simeq 2\,{\rm eV}$. This means that for $m_1\gtrsim
2\,{\rm eV}$ the fully flavored behavior does not obtain. Notice
also that this upper limit is quite conservative because the lower bound
on $M_1$ has been obtained neglecting that, for $K_1=m_1/m_{\star}$,
the flavored {\it CP\/} asymmetry vanishes and thus the bound cannot
be saturated. Moreover we have assumed that $P^0_{1\a}$ can always
assume the value that maximizes the asymmetry.

In Fig.~\ref{fig:mMplot} we also show (thin solid line) a lower
bound $M_1(m_1)$ in a specific scenario~\cite{Blanchet:2006be},
corresponding to a particular choice of the orthogonal see-saw
matrix~\cite{Casas:2001sr} $\O=R_{13}$, that represents a complex
rotation in the 13-plane, where $\O^2_{13}$ is taken purely imaginary.
In this case the value of $z_{B\a}$ is not necessarily the same
that maximizes the asymmetry in the one-flavor dominated scenario and
$K_1> m_1/m_{\star}$. Therefore,
a specific calculation  is  necessary in order to work out correctly
the condition Eq.~(\ref{M1fl}).
The result is shown in Fig.~\ref{fig:mMplot} with a thin dashed line.

In this case the upper limit on $m_1$ for
the applicability of the fully flavored regime is much smaller,
$m_1\simeq 0.1\,{\rm eV}$. Allowing for a non-vanishing
real part of $\O^2_{13}$, slightly larger values are possible.
It should be however kept in mind that these values are indicative since they
rely on a condition for the fully flavored regime that comes from a
simple rate comparison.

%%%%%%%%%%%%%%%%%%%%%%%%%%%%%%%%%%%%%%%%%%%%%%%%%%%%%%%%%%%%%%%%%%%%%%
\section{Limitations of a simple rate comparison}   \label{sec:beyond}
%%%%%%%%%%%%%%%%%%%%%%%%%%%%%%%%%%%%%%%%%%%%%%%%%%%%%%%%%%%%%%%%%%%%%%

We have exploited a somewhat qualitative rate comparison for the
determination of the region where the fully flavored regime obtains.
While be believe that our approach nicely illustrates the
modifications that derive from our more restrictive criterion for
the significance of flavor effects, there are also important
shortcomings. First, we have simply compared the inverse decay rate
with the charged-lepton Yukawa interaction rate, ignoring flavor
oscillations caused by the flavor-dependent lepton dispersion
relation in the medium. If the oscillations are much faster than the
inverse decay rate, they also contribute effectively, together with
inelastic scatterings, to project the lepton state on the flavor
basis. Therefore, including oscillations will tend to enlarge the
region where the fully flavored behavior obtains (the inclined upper
dot-dashed line in Fig.~\ref{fig:mMplot}) and to reduce that one
where the unflavored behavior obtains (the horizontal upper
dot-dashed line in Fig.~\ref{fig:mMplot}). In our case, the
oscillation frequency is comparable to $\G_{\alpha}$ and so the two
estimations are not too far off.

Moreover, we have also neglected $\D L=1$ scatterings. They also
contribute, like inverse decays, both to generate the asymmetry and
to the wash out and therefore, together with inverse decays,
contribute to preserving the flavor direction of the leptons. At the
relevant $z\sim z_{B\a}\sim 2$, the $\D L=1$ scattering rate is
actually larger than the inverse decay rate and thus tends to reduce
the region where the fully flavored behavior obtains
(the lower inclined dot-dashed line in Fig.~\ref{fig:mMplot}). Therefore, the
effects of oscillations and of $\D L=1$ scatterings may partially
cancel each other. In Fig.~\ref{fig:mMplot} the region between the
two inclined sot-dashed lines
gives therefore an indication of the theoretical uncertainty on the
determination of the region where the fully flavored regime
holds. It can be seen that current calculations cannot establish whether
the upper bound holding in the unflavored regime is
nullified, just simply relaxed or still holding, when flavor effects
are included.

Only a full quantum-kinetic treatment can give a final verdict on
the effectiveness of flavor effects in leptogenesis and its impact
on the neutrino mass limit. While we have verified that our rate
criteria are borne out by the quantum kinetic equations stated in
Ref.~\cite{Abada:2006fw}, these equations are not necessarily
complete in that the terms describing the generation and wash out of
the asymmetry have been added by hand. Moreover, the ``damping
rate'' caused by the flavor-sensitive Yukawa interactions ultimately
derives from a collision term in the kinetic
equation~\cite{Raffelt:1992uj}. Extending the pioneering treatment
of Ref.~\cite{Abada:2006fw} to allow for a complete understanding of
flavor effects remains a challenging~task.

%%%%%%%%%%%%%%%%%%%%%%%%%%%%%%%%%%%%%%%%%%%%%%%%%%%%%%%%%%%%%%%%%%%%%%
\section{Conclusions}                          \label{sec:conclusions}
%%%%%%%%%%%%%%%%%%%%%%%%%%%%%%%%%%%%%%%%%%%%%%%%%%%%%%%%%%%%%%%%%%%%%%

Flavor effects can play a very important role in leptogenesis.
However, we have shown that the condition for the fully flavored
behavior is more restrictive because one needs to compare the speed
of the charged-lepton Yukawa interactions with that of
$N\leftrightarrow\ell+\Phi$, not with the cosmic expansion rate.
This distinction makes a significant difference particularly in the
strong wash-out regime where for some of the leptogenesis epoch the
rate of $N\leftrightarrow\ell+\Phi$ is faster than the cosmic
expansion.

We are especially interested in the question if the traditional
neutrino mass bound of the unflavored
treatment~\cite{Buchmuller:2003gz} can be circumvented by flavor
effects. We have found that the see-saw parameters that correspond
to $m_1\gtrsim 0.1\,{\rm eV}$ do not fall into the strictly
unflavored or the fully flavored regimes. The intermediate regime
requires a detailed quantum kinetic treatment, so at present it is
not possible to decide if the upper bound neutrino mass bound can be
circumvented (relaxed or nullified) by flavor effects.

%%%%%%%%%%%%%%%%%%%%%%%%%%%%%%%%%%%%%%%%%%%%%%%%%%%%%%%%%%%%%%%%%%%%%%
\section*{Acknowledgments} %%%%%%%%%%%%%%%%%%%%%%%%%%%%%%%%%%%%%%%%%%%
%%%%%%%%%%%%%%%%%%%%%%%%%%%%%%%%%%%%%%%%%%%%%%%%%%%%%%%%%%%%%%%%%%%%%%

It is a pleasure to thank W. Buchm\"{u}ller, S. Davidson, M. Losada,
M. Pl\"{u}macher and E. Roulet for comments and discussions.
We also wish to thank A. Riotto for sending us the draft of a forthcoming
manuscript with A. De Simone on the same subject.
This work was supported, in part, by the Deutsche
Forschungsgemeinschaft (DFG) under grant No.~SFB-375 and by the
European Union under the ILIAS project, contract
No.~RII3-CT-2004-506222, and under the Marie Curie project
``Leptogenesis, Seesaw and GUTs,'' contract No.~MEIF-CT-2006-022950.

\newpage

%%%%%%%%%%%%%%%%%%%%%%%%%%%%%%%%%%%%%%%%%%%%%%%%%%%%%%%%%%%%%%%%%%%%%%
\section*{References} %%%%%%%%%%%%%%%%%%%%%%%%%%%%%%%%%%%%%%%%%%%%%%%%
%%%%%%%%%%%%%%%%%%%%%%%%%%%%%%%%%%%%%%%%%%%%%%%%%%%%%%%%%%%%%%%%%%%%%%

%%%%%%%%%%%%%%%%%%%%%%%%%%%%%%%%%%%%%%%%%%%%%%%%%%%%%%%%%%%%%%%%%%%%%%
\end{document}